# Soliton generation in CaF$_2$ crystalline whispering gallery mode resonators with negative thermal-optical effects

Mingfei Qu,[1,2] Chenhong Li,[1,2] Kangqi Liu,[1,2] Weihang Zhu,[1,2] Yuan Wei,[1] Pengfei Wang,[1] Songbai Kang[1,*]

[1]Key Laboratory of Atomic Frequency Standards, Innovation Academy for Precision Measurement Science and Technology, Chinese Academy of Sciences, Wuhan 430071, China
[2]University of Chinese Academy of Sciences, Beijing 100049, China
*Corresponding author: kangsongbai@apm.ac.cn

Calcium fluoride (CaF$_2$) crystalline whispering gallery mode resonators (WGMRs) exhibit ultrahigh intrinsic quality factors and a low power anomalous dispersion in the communication and mid-infrared bands, making them attractive platforms for microresonator-based comb generation. However, their unique negative thermo-optic effects pose challenges when achieving thermal equilibrium. To our knowledge, our experiments serve as the first demonstration of soliton microcombs in Q > 10$^9$ CaF$_2$ WGMRs. We observed soliton mode-locking and bidirectional switching of soliton numbers caused by the negative thermo-optic effects. Additionally, various soliton formation dynamics are shown, including breathing and vibrational solitons, which can be attributed to thermo-photomechanical oscillations. Thus, our results enrich the soliton generation platform and provide a reference for generating solitons from WGMRs that comprise other materials with negative thermo-optic effects. In the future, the ultrahigh quality factor of CaF$_2$ crystal cavities may enable the generation of sub-milliwatt-level broad-spectrum soliton combs.

With the rapid advancement of micro- and nano-fabrication techniques, optical microresonators have become revolutionary devices characterized by their extremely low power thresholds for nonlinear effects, which are a result of their ultrahigh quality factors and unprecedentedly small mode volumes [1,2]. A Kerr frequency comb [3,4] has been observed based on four-wave mixing (FWM) caused by the third-order Kerr effect in a microresonator. Recently, a temporal dissipative Kerr soliton (DKS) has enabled the formation of a stable mode-locking Kerr microcomb. A DKS is a sequence of stable time-domain soliton pulses achieved through the dual balance of dispersion and the Kerr effect, as well as parametric gain and linear loss [5]. DKSs are employed in many applications such as in optical clocks [6], coherent communications [7], ultrastable microwave generation [8], optical frequency synthesis [9], spectroscopy [10], and LIDAR [11].

Currently, a range of materials are utilized for soliton microcomb generation, including MgF$_2$ [12,13], SiO$_2$ [14], Si$_3$N$_4$ [15], Al$_2$N$_3$ [16,17], LiNbO$_3$ [18], Si [19], Hydex [20], and AlGaAs [21]. Most of these are on-chip resonators that are compatible with microelectromechanical system fabrication. While on-chip microresonators are compact and effective platforms for Kerr soliton generation, mechanically polished fluoride crystalline whispering gallery mode resonators (WGMRs) such as MgF$_2$, CaF$_2$, and BaF$_2$ are interesting alternatives that receive extensive attention owing to their ultrahigh quality factor (Q > 10$^9$) and flat anomalous group velocity dispersion region from communication to the mid-infrared[22]. However, only MgF$_2$ crystalline WGMRs have been widely used to generate DKSs [12,13].

Significant progress has been made in using CaF$_2$ WGMRs as experimental platforms for producing microcombs. Theoretical predictions suggest that the quality (Q) factor limit of a CaF$_2$ WGMR can reach 10$^{14}$ at room temperature [23]. Experiments utilizing mechanical polishing combined with an annealing process have demonstrated Q-factors greater than 10$^{11}$ for this material [24-26]. CaF$_2$ has the potential to be an ultralow-power consumption (submilliwatt) platform for producing microcombs; however, its negative thermal-optic (TO) effects complicate the thermal equilibrium process. To date, the generation of solitons in CaF$_2$ crystalline WGMRs has not yet been successfully demonstrated [27,28]. In this letter, we report the successful generation of solitons in homemade CaF$_2$ WGMRs that exhibit self-starting and bidirectional switching of their soliton states. Additionally, we observed several soliton formation dynamics such as breathing and vibrational solitons

[29,30], which are believed to be associated with thermophotomechanical oscillations.

To demonstrate soliton generation, a platform that had a Q factor of $1.5\times10^9$ was constructed from a $CaF_2$ crystal using homemade a hand-polishing technique developed by these authors, as shown in Fig. 1(a). The power evolution of the pump laser was measured experimentally during the resonance frequency tuning process, as shown in Fig. 1(b). The negative thermal effect of $CaF_2$ leads to a characteristic hysteretic response, resulting in a drift toward higher frequencies, irrespective of whether the laser frequency creates a forward or backward scan. These thermal dynamics can lead to thermal cavity self-locking under pump-resonator red detuning [27].

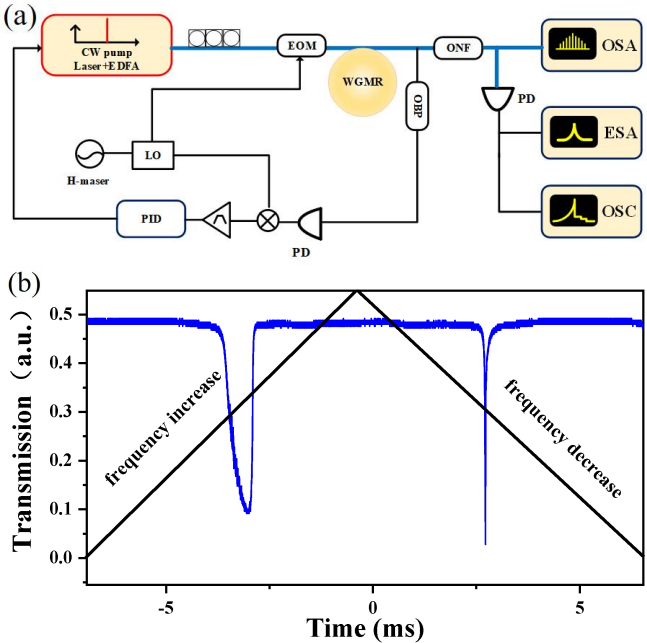

Fig. 1. Experimental platform utilized for soliton generation. (a) Key elements of the setup, including the fiber polarization control (FPC), electro-optic phase modulator (EOM), erbium-doped fiber amplifier (EDFA), local oscillator (LO), optical notch filter (ONF), optical band pass (OBP), photoelectric detector (PD), oscilloscope (OSC), electrical spectrum analyzer (ESA), and optical spectrum analyzer (OSA). (b) Measured transmission of the pump power during up and down wavelength scans of the laser, which show an inverted triangle of power as the frequency increases (red to blue) due to the negative thermo-optic effects.

Generating stable solitons in microresonators based on WGM poses a significant challenge owing to the ultrasmall effective mode area of these structures, which induces a strong thermal-optical effect within the cavity. Furthermore, the $CaF_2$ WGMR experiences thermo-opto-mechanical oscillations owing to competition between the negative TO effect and thermal expansion (TE), leading to a complex progress that results in a stable soliton state. To initiate the soliton, the pump frequency was scanned from blue- to red-detuned, and the power levels of the intra-cavity microcomb were measured, which are shown in Fig. 2. Notably, unlike platforms that rely on positive thermal-optical effects, the $CaF_2$ WGMR lacks a modulation-instability comb in blue-detuned regions, with soliton steps emerging only in red-detuned regions. To mitigate this, we utilized the sideband Pound–Drever–Hall (PDH) method to actively restrain the variations in pump resonance detuning due to thermal-optomechanical oscillation, thereby ensuring stable single- or multi-soliton state(s).

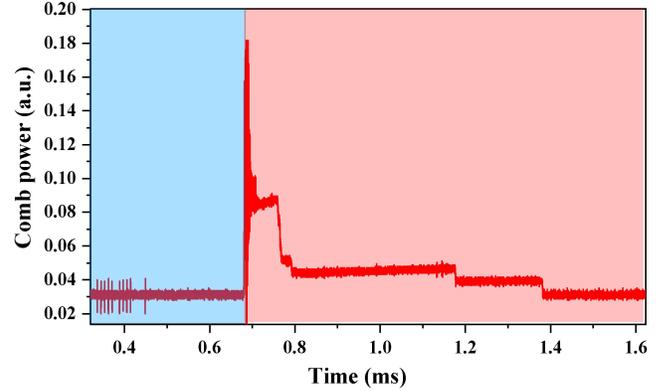

Fig. 2. Generation of Kerr solitons in a $CaF_2$ WGMR. The level of intra-cavity comb power corresponds to the laser frequency scanning from blue-detuned to red-detuned, with the red and blue shadows indicating the approximate red-detuned and blue-detuned regions, respectively.

The laser frequency was capped at the power level of a single soliton, and the spectral information was detected using a spectrometer. Figure 3 shows the optical spectrum of a single soliton, which exhibits a 2-nm red shift from the pump frequency and a strong asymmetry. This redshift results from soliton recoil owing to the emission of stronger dispersive waves on the high-frequency side caused by Cherenkov radiation. Moreover, dispersive wave generation was enabled by avoided mode crossing owing to the dense modes of the $CaF_2$ microresonator. Additionally, the Raman effect is negligible in millimeter-sized $CaF_2$ crystalline resonators owing to their narrow gain bandwidths [31]. Further analysis of the dispersion waves is presented in Supplement 1.

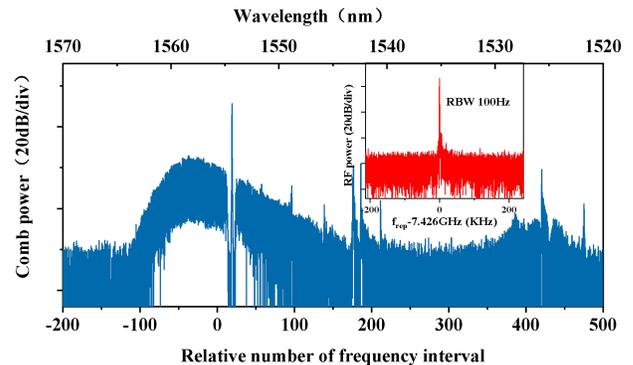

Fig. 3. Optical spectrum of a single soliton comb with 7.426 GHz comb spacing. The inset is a repetition frequency ($f_{rep}$ = -7.42 GHz) RF signal of a soliton comb (RBW 100 Hz).

To demonstrate soliton coherence, a 20-GHz bandwidth photodetector was utilized to detect the RF beat note signal of the soliton comb. The 7.426 GHz RF signal (mode spacing matches the resonator FSR) has a signal-to-noise ratio of over 50 dB, indicating high coherence among the soliton comb teeth, as shown in the inset of Figure 3.

The negative TO effect of the $CaF_2$ WGMRs shifts the resonator wavelength in the opposite direction of the shift caused by the Kerr effect and TE, similar to the photorefractive effect of $LiNbO_3$ microresonators [18]. This property enables the self-stabilization of the laser cavity frequency detuning around the red-detuned regime. To investigate this phenomenon, the laser frequency was scanned up and down on the red-detuned side of the cavity mode at approximately 1556 nm. Figure 4 shows the intra-cavity power as a function of the scan time, where time zero indicates the scanning corner point. These results revealed that a complex dynamic process leads to soliton formation, involving chaotic modulation combs (Regions II, IV), vibrational solitons (Region III) [29,30], breathing solitons (Region V), and stable solitons (Region VI).

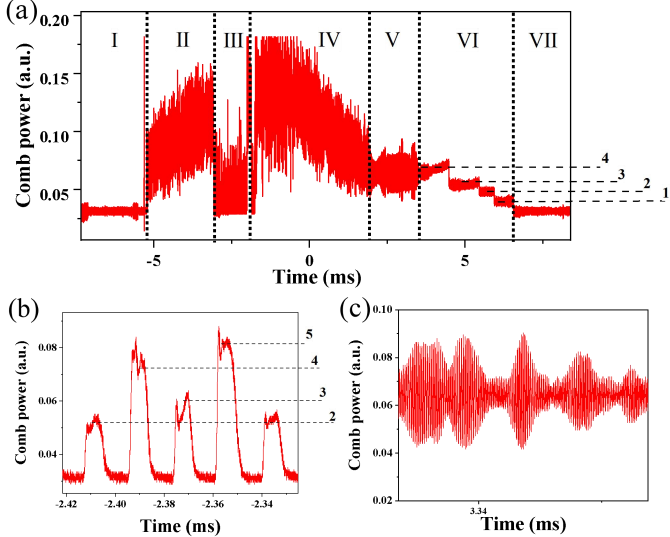

Fig. 4. Deterministic Kerr soliton generation and solitons in different states. (a) Comb power as a function of time while scanning the laser frequency in the red-detuned region in forward and backward directions. The power step represents the soliton triggering, and the dashed line represents the number of solitons. Panels (b) and (c) show the temporally resolved waveforms of Regions III and V, respectively, with the dashed line denoting the number of solitons; (b) illustrates the periodic excitation and annihilation of local solitons, and (c) depicts oscillations in the generated comb power.

When the laser frequency approaches the resonant frequency and reaches the gain threshold of the modulation instability, the comb power increases, indicating that the initial continuous wave field evolves into chaotic modulation combs (Region II). Of particular interest is Region III where the comb power drops instantaneously, indicating a different soliton formation state than one resulting from typical dynamics. Fig. 4(b) magnifies part of Region III, showing localized wave packets in the time domain similar to those in a photomechanical resonator [29]. Dashed lines indicate the number of solitons corresponding to those shown in Fig. 4(a). This phenomenon was caused by the thermo-opto-mechanical oscillation of the $CaF_2$ microresonator. As the laser frequency decreases, the comb power diminishes discretely, indicating a transition between the soliton states. In our results, the transition into the four-breathing soliton regime coincides with the power oscillation shown in Fig. 4(c), which then gradually transitions into the stable soliton regime. Remarkably, we observed that a decrease in laser frequency deterministically produced a single soliton. The backward tuning process results in the continuous extinction of the soliton (i.e., the soliton switch) until a single soliton state is reached (N → N-1, …, 1), as shown in Figure 4.

Currently, scanning frequency from the blue- to the red-detuned region is necessary to trigger soliton states and achieve thermal equilibrium. As shown in the aforementioned experimental results, soliton excitation can be achieved without inducing high-power chaotic states owing to the negative TO effect. However, the soliton switches are unidirectional in most experimental systems. Bidirectional switching between different soliton states has recently been reported in $LiNbO_3$ systems [18] that employ the photorefractive effect and thermal overcompensation of an auxiliary laser system [32]. This phenomenon may also be induced in $CaF_2$ crystalline WGMRs, where the negative TO effect causes the resonance shift to be inversely proportional to the power variation.

We provide a detailed description of the variation in the cavity resonance pump detuning during the bidirectional switching of different soliton states in Figure 5. The intra-cavity power (indicated in red) and error signal (indicated in black) of the PDH were recorded after the pump laser swept across the resonance from the blue side. The soliton states were ignited by tuning the pump laser frequency and stopping it at the time = 0 ms, as shown in Figure 5.

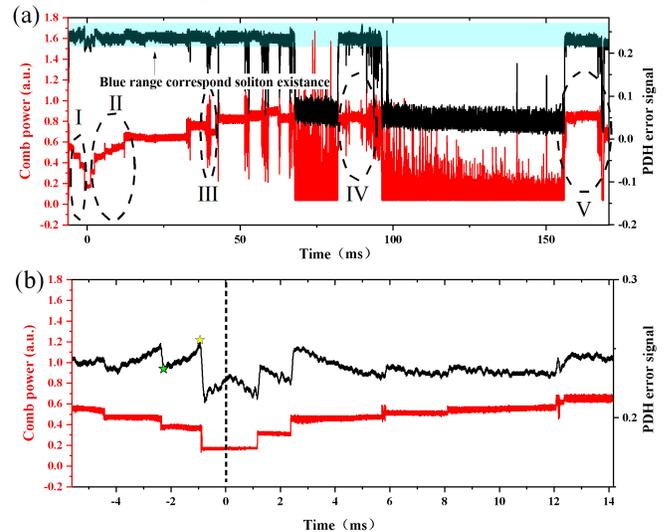

Fig. 5. Bidirectional switching of the soliton states. Panel (a) displays the experimental trace of the comb power and the corresponding Pound–Drever–Hall (PDH) error signal, which depict the change in the amount of detuning during backward-tuned pump laser usage (time<0) and after scanning stops (time>0). Dashed black circles marked with I, II, III, IV, and V indicate the regions associated with decreasing soliton numbers, soliton bursts, power spikes, and soliton self-start processes. The blue region shows the pump-resonator detuning when the soliton is generated. Panel (b) amplifies the region near the stop of the pump optical frequency scanning for better detuning visualization.

During the sweep phase, the laser frequency decreased more than the resonance redshift as a result of the negative TO effect following from the reduced power. This caused a gradual increase in the effective resonance pump detuning, eventually reaching the maximum range suitable for the soliton, where it was lost. Subsequently, the system entered the next soliton state, and the intra-cavity power experienced a sharp decline. The Kerr effect is attributed to this rapid drop in the intra-cavity power, which caused a sudden blueshift in the resonance wavelength. The effective detuning quickly exceeded the soliton existence range. Subsequently, a gradual redshift of the resonance wavelength occurred because of the negative TO effect. Fig. 5(a) and (b) show the intra-cavity power and corresponding detuning changes, respectively. The minimum and maximum effective detuning amounts for the soliton existence are represented in Fig. 5(b) by five green and yellow stars, respectively, corresponding to the presence of 2-soliton states.

We emphasize that with this mechanism, the laser frequency need not be swept to ignite the soliton states. Laser scanning was halted at time = 0 ms, leading to a redshifted resonance frequency owing to the negative TO effect of the cavity. The effective detuning gradually decreases to the region of soliton existence and transiently promotes a detuning regime where instabilities exist, leading to the onset of a power spike (region III in Fig. 5(a)). A significant change in the amount of detuning occurred when the power spiked from the PDH error signal. When the soliton existence state was entered again, a soliton state with a certain number of N solitons (determined by the cavity lifetime) was obtained at a fast rate [18]. The soliton reappeared spontaneously near 80 and 160 ms after the loss, demonstrating its self-starting ability. Upon self-starting the soliton, the PDH error signal indicates the re-entry of the detuning amount into a range suitable for the soliton.

In summary, we have demonstrated the generation of solitons in $CaF_2$ crystal WGMRs with negative TO effects. We achieved the long-term trapping of solitons using the offset PDH technique. Our experiments revealed a rich soliton formation dynamic process, including a multi-soliton state, single soliton state, breathing solitons, and similar vibrational solitons due to photomechanical oscillation. $CaF_2$ has a negative TO effect that can facilitate soliton-mode-locked self-starting and the mutual bidirectional switching of soliton states. Our experimental results enrich the platform for dissipative soliton microcombs and provide a reference for generating solitons from WGMRs composed of other materials with negative TO effects.

**Funding.**

**Acknowledgments**. We thank Prof J. XS helpful discussion on the paper.

**Disclosures.** The authors declare no conflicts of interest.

**Data availability.** The data underlying the results presented in this paper are not publicly available at this time but may be obtained from the authors upon reasonable request.

**Supplemental document.** See Supplement 1 for supporting content.